\documentclass[byrevtex,prb,aps,nobibnotes,twocolumn]{revtex4}
\usepackage{graphicx}%
\usepackage{dcolumn}
\usepackage{amsmath}   
\usepackage{bm}
\usepackage{dcolumn}
\usepackage{longtable}
\begin{document}

\title{\centering\Large\bf Reorganization asymmetry of electron transfer in 
ferroelectric media and principles of artificial photosynthesis }
\author{Dmitry V.\ Matyushov} 
\email[E-mail:]{dmitrym@asu.edu.}  
\affiliation{
  Department of Chemistry and Biochemistry and the Center for the
  Early Events in Photosynthesis, Arizona State University, PO Box
  871604, Tempe, AZ 85287-1604}
\date{\today}
\begin{abstract}
  This study considers electronic transitions within donor-acceptor
  complexes dissolved in media with macroscopic polarization.  The
  change of the polarizability of the donor-acceptor complex in the
  course of electronic transition couples to the reaction field of the
  polar environment and the electric field created by the macroscopic
  polarization. An analytical theory developed to describe this
  situation predicts a significant asymmetry of the reorganization
  energy between charge separation and charge recombination
  transitions. This result is proved by Monte Carlo simulations of a
  model polarizable diatomic dissolved in a ferroelectric fluid of
  soft dipolar spheres. The ratio of the reorganization energies for
  the forward and backward reactions up to a factor of 25 is
  obtained in the simulations. This result, as well as the effect of
  the macroscopic electric field, is discussed in application to the
  design of efficient photosynthetic devices.
\end{abstract}
\maketitle

\section{Introduction}
\label{sec:1}
\begin{figure}[htbp]
  \centering \includegraphics*[width=6cm]{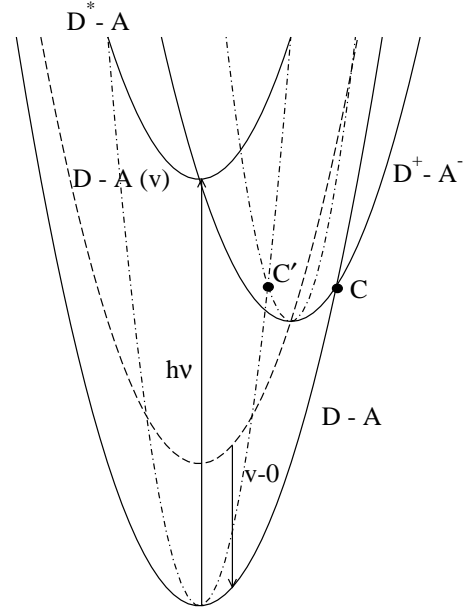}
  \caption{Energetics of photoinduced electron transfer involved in
    natural and artificial photosynthesis.  The charge-separated state
    D$^+$--A$^-$ is placed below the photo-excited state D$^*$--A by
    the reorganization energy $\lambda$ to ensure activationless transition.
    The dashed line indicates the vibrationally excited electronic
    ground state for which the recombination transition
    D$^+$--A$^-\to$D--A does not require an activation barrier.  The
    dash-dotted lines mark the free energy surfaces of states
    D$^+$--A$^-$ and D--A with the reorganization energy four times
    lower than the value of $\lambda$ used to draw the free energy surfaces
    shown by the solid lines.  C and C$'$ indicate the classical
    transition states. }
  \label{fig:1}
\end{figure}

Electron transfer (ET) reaction is a basic elementary step in
photoinduced charge separation occurring in both natural and
artificial photosynthesis. The fundamental mechanism behind the
conversion of light energy into the energy of a charge-separated state
is illustrated in Figure \ref{fig:1}.  Optical excitation $h\nu$ of the
donor unit, D$\to$D$^*$, within the donor-acceptor complex, D--A, lifts
the energy level of the donor to create conditions for the
photoinduced charge separation step, D$^*$--A $\to$ D$^+$--A$^-$.
Charge separation is generally an activated transition.  The
activation barrier, according to the Marcus-Hush
theory,\cite{Marcus:93} can be obtained from the equilibrium free
energy gap between the final and initial ET states, $\Delta F_0$, and the
free energy $\lambda$ required to reorganize the nuclear degrees of freedom
(reorganization free energy). When the equilibrium free energy of the
acceptor is below the free energy of the donor by the amount of free
energy $\lambda$, the charge separation transition is activationless.  Such
an activationless transition is often fast and efficient, but it
requires loosing the energy $\lambda$ from the photon energy $h\nu$.
Increasing the energetic efficiency of photosynthesis therefore demands
$\lambda$ to be as low as possible. It is often assumed that an important
role played by the hydrophobic protein matrix in facilitating ET in
natural photosynthesis is to screen highly polar aqueous environment
and to reduce $\lambda$.

Once the charge-separated state has been created, it needs to be
sufficiently long-lived to be used for energy storage. A mechanism
often suggested to be of primary importance in natural systems is the
use of a sequence of activationless ET steps to move the electron away
from the primary donor. Each activationless step $i$ lowers the system
energy by $\lambda_i$ thus resulting in the overall low energetic efficiency
of the photosynthetic unit. The reaction competing with photoinduced
charge separation is the normally highly exothermic return electron
transfer to the ground D--A state. The classical reaction path then
goes through the relatively high activation barrier in the inverted ET
region predicted by the Marcus-Hush theory (point ``C'' in Figure
\ref{fig:1}).\cite{Marcus:93} However, the system avoids this
activated path by nearly activationless transfer to a vibrationally
excited state, D--A(v), of the electronically ground donor-acceptor
complex (dashed line in Figure \ref{fig:1}).  The vibrational energy
is subsequently released to heat through vibrational relaxation ($v\to
0$ in Figure \ref{fig:1}).

The design an optimization of a sequence of activationless ET
reactions requires a very precise molecular tuning, which is hard to
achieve in synthetic systems despite some significant progress
achieved in this field in recent years.\cite{Wasielewski:92,Moore:93}
It is therefore desirable to search for mechanisms of reducing the
rate of return ET in molecular photosynthesis with the goal of
achieving higher quantum yield for the charge-separated state.
According to the current understanding of radiationless transitions in
molecules,\cite{BixonJortner:99} an efficient way to reduce the return
rate would be to move the reaction into the normal region of ET.

The electronic states D$^*$--A and D--A are distinct and, in
principle, the charge-recombination transition D$^+$--A$^- \to$ D--A can
be characterized by a pair of parabolas with the curvatures different
from those of the charge-separation transition D$^*$--A $\to$
D$^+$--A$^-$.  In the reactions classification adopted here we label
transition D$^*$--A $\to$ D$^+$--A$^-$ as charge separation and
transition D$^+$--A$^-$ $\to$ D--A as charge recombination. Backward
transitions at each step are not considered as separate steps in the
reaction mechanism.  When the photoexcitation energy is kept constant,
shifting the recombination reaction into the normal region would
require increasing the curvature of the charge-recombination parabolas
(dash-dotted lines in Figure \ref{fig:1}), i.e.\ lowering the
reorganization energy.  The inverted-region activated state C then
shifts to the normal-region activated state C$'$ in Figure
\ref{fig:1}.

It is easy to realize the pitfall of the picture shown in Figure
\ref{fig:1}.  The separation of the minima of two dash-dotted
parabolas must be equal, in the Marcus-Hush theory, to twice the
reorganization energy, which is clearly violated once the curvature is
increased without corresponding shift of the parabolas.  Therefore,
the goal of bringing the recombination reaction to the normal region
can be realized only within models extending beyond the Marcus-Hush
picture of equal-curvature parabolas. On needs flexibility, built into
the model, that would allow decoupling of the Stokes shift from the
curvatures. The use of parabolic free energy surfaces with different
curvatures, as was suggested by Kakitani and Mataga,\cite{Kakitani:85}
is prohibited by the requirement of energy conservation. When the
energy gap between the donor and acceptor energy levels is taken for
the reaction coordinate, the free energy surfaces of the initial,
$F_1(X)$, and final, $F_2(X)$, ET states are connected by the linear
relation established by Warshel\cite{Hwang:87} and
Tachiya\cite{Tachiya:89}
\begin{equation}
  \label{eq:1-1}
  F_2(X) = F_1(X) + X 
\end{equation}
The parabolic surfaces with different curvatures clearly violate this
requirement and, therefore, cannot be used for the modeling of ET
reactions. The problem can be resolved within a three-parameter model
of ET free energy surfaces\cite{DMjcp:00} which allows different
reorganization energies and, at the same time, does not violate eq
\ref{eq:1-1}. The free energy surfaces $F_i(X)$ are then necessarily
non-parabolic.

Once different reorganization energies are allowed within a theory of
free energy surfaces, one needs to address the following major
question: What would be a mechanism resulting in dramatic changes of
reorganization energies between charge separation and charge
recombination reactions? A question relevant to studies of natural
photosynthesis is the following: Is the role of the protein matrix can
be reduced to providing a low dielectric constant or there are some
other properties of proteins beneficial for high efficiency of natural
photosynthesis? Note that shifting the transition point from C to C$'$
requires a very significant change in the reorganization energy (four
times decrease in Figure \ref{fig:1}). The possibility of different
equilibrium-point curvatures of the two ET free energy surfaces has
been actively pursued in the last decades by computer simulations
searching for the effects of non-linear solvation on
ET.\cite{Hwang:87,Kuharski:88,Marchi:93,Yelle:97,Hartnig:01,DMjcp2:04}
However, essentially in all reports, the difference in reorganization
energies between the initial and final ET states has never come close
to the magnitude that would dramatically change the mechanism of
charge recombination.

In this paper, we address the problem of reorganization asymmetry by
considering ET reactions in polarizable donor-acceptor complexes
immersed in polarization anisotropic media. A part of the motivation
for this approach comes from studies of bacterial photosynthesis. The
protein matrix in natural bacterial reaction centers is highly
anisotropic with the electric field amounting $\simeq 10^7$
V/cm.\cite{Steffen:94} On the other hand, the primary pair is highly
polarizable with the polarizability change upon photoexcitation about
800--1100 \AA$^3$.\cite{Middendorf:93} Since bacteriochlorophyll
cofactors are much less polarizable,\cite{Kjellberg:03} primary charge
separation occurs with a large negative polarizability change.  On a
more fundamental level, polarizability change in the course of ET
leads to very significant reorganization asymmetry, as follows from
both analytical theories\cite{DMjpca:99} and computer
experiment.\cite{DMjacs:03,DMjpca:04} The asymmetry arising from the
coupling of changing polarizability to the solvent recation field is
by far the largest achieved so far in computer simulations thus
promising the desired alteration of the energetics of photosynthetic
reactions.

In Section \ref{sec:2}, we present an analytical model of ET free
energy surfaces when both the polarizability and the dipole moment of
the donor-acceptor complex change in the course of ET.  The
donor-acceptor complex is immersed in a polar medium which, in
addition to usual polar response, is characterized by a macroscopic
electric field. In order to mimic high electric fields present in
molecular systems (thin films, proteins, etc.), we use a polar model
fluid capable of producing ferroelectric order (Section \ref{sec:3}).
The ordered phase is obtained by Monte Carlo (MC) simulations.  The
possibility of ferroelectric order in bulk polar fluids is still
actively debated in the literature,\cite{Wei:92,Zhang:95,Morozov:03}
and it has been suggested that boundary conditions employed in the
simulation protocol significantly affect the possibility of creation
of the macroscopically polar phase.\cite{Wei:93} For our current
purpose, this subject is not relevant since we are mostly interested
in reactions in condensed phases with macroscopic electric fields
strong enough to influence properties of electronic transitions.
Whether the macroscopic polar order is stabilized by surface effects
or some other reasons is not central to our study.  On the other hand,
ferroelectrics and ordered polar films may be used as solvents for ET
reactions. The present analysis then directly applies to such systems.
We return to discussing the role of reorganization assymetry in
improving the efficiency of phtosynthesis in Section \ref{sec:4}.

\section{Energetics of electron transfer}
\label{sec:2}
\subsection{Model}
\label{sec:2-1}
Consider a donor-acceptor complex immersed in a medium with a nonzero
macroscopic electric field $\mathbf{F}$. The complex has the dipole moments
$\mathbf{m}_{0i}$ and polarizabilities $\bm{\alpha}_{0i}$ in two electronic
states, $i=1,2$. These states are coupled to nuclear fluctuations in
the solvent and are ``dressed'' with the field of the electronic
solvent polarization following adiabatically the changes in the charge
distribution within the complex. Because of the electronic
polarization, the dipole moments and polarizabilities are different
from their gas phase values $\mathbf{m}_{0i}$ and
$\bm{\alpha}_{0i}$:\cite{DMjpca:99}
\begin{equation}
  \label{eq:2-1}
  \mathbf{m}_i = \left[\mathbf{1}-2\bm{\alpha}_{0i}\cdot\mathbf{a}_e \right]^{-1}\cdot \mathbf{m}_{0i}
\end{equation}
and 
\begin{equation}
  \label{eq:2-2}
  \bm{\alpha}_i = \left[\mathbf{1}-2\bm{\alpha}_{0i}\cdot\mathbf{a}_e \right]^{-1}\cdot \bm{\alpha}_{0i}
\end{equation}
Here, $\mathbf{a}_e$ is the linear response function such that the
free energy of solvation of dipole $\mathbf{m}_0$ by the electronic
solvent polarization (subscript ``e'') is
$\mathbf{m}_0\cdot\mathbf{a}_e\cdot\mathbf{m}_0$. Once the electronic degrees
of freedom have been adiabatically eliminated, the electronic energy
levels of the charge-transfer complex are affected by the microscopic
nuclear field $\mathbf{R}_n$ (nuclear reaction field, subscript ``n'') and the
macroscopic field $\mathbf{F}$ of the solvent:\cite{DMjpca:99}
\begin{equation}
  \label{eq:2-3}
  E_i[\mathbf{R}_n] = I_i^{\text{np}} - \mathbf{m}_{i}\cdot(\mathbf{R}_n+\mathbf{F}) - 
        \frac{1}{2}(\mathbf{R}_n+\mathbf{F})\cdot\bm{\alpha}_{i}\cdot (\mathbf{R}_n +\mathbf{F}) 
\end{equation}
Here $I_i^{\text{np}}$ are the energies of the ET complex which
include the gas-phase energies and free energies of solvation by the
solvent electronic polarization expressed as a sum of induction and
dispersion solvation components.

The system Hamiltonian 
\begin{equation}
  \label{eq:2-4}
  H_i[\mathbf{R}_n] = E_i[\mathbf{R}_n] + H_B[\mathbf{R}_n]
\end{equation}
is the sum of energies $E_i[\mathbf{R}_n]$ and the solvent bath
Hamiltonian $H_B$. The Gaussian (linear response) approximation is
adopted for the latter
\begin{equation}
  \label{eq:2-5}
  H_B[\mathbf{R}_n] = \frac{1}{4} \mathbf{R}_n\cdot\mathbf{a}_n^{-1} \cdot \mathbf{R}_n 
\end{equation}
where $\mathbf{a}_n$ is the linear response function of the solvent
nuclear polarization. In the following, for simplicity, we will assume
collinear dipole moments $\mathbf{m}_i$ and the polarizability
changing its value only along the direction of the dipole moment. We
will also assume that the solute polarizability increases for the $1\to
2$ transition, i.e., $\Delta\alpha=\alpha_2-\alpha_1>0$. These approximations allow us to
consider $\mathbf{a}_n$ as a scalar with the only non-zero projection
along the solute dipole. We will also define the scalar $F_m$ as the
projection of the external field $\mathbf{F}$ on the direction of the
solute dipole moment. The consideration of a more general case does
not present fundamental difficulties.\cite{DMjpca:99}

\subsection{Free energy surfaces}
\label{sec:2-2}
The classical Hamiltonian $H_i[\mathbf{R}_n]$ in eq \ref{eq:2-4} can
be used to build the free energy surfaces of ET along the reaction
coordinate associated with the fluctuating donor-acceptor energy gap
required to reorganize the nuclear degrees of freedom 
\begin{equation}
  \label{eq:2-6}
  X =\Delta H[\mathbf{R}_n] = H_2[\mathbf{R}_n] - H_1[\mathbf{R}_n]
\end{equation}
The free energy surfaces for the initial ($i=1$) and final ($i=2$) ET states are 
obtained by constrained integration over the nuclear reaction field
\begin{equation}
  \label{eq:2-7}
  e^{-\beta F_i(X)} = A \int \delta(X- \Delta H[\mathbf{R}_n]) e^{-\beta H_i[\mathbf{R}_n]} d\mathbf{R}_n ,
\end{equation}
where $A$ is used to account for the units of the field $\mathbf{R}_n$ and
$\beta=1/ k_{\text{B}}T$. 

\begin{figure}[htbp]
  \centering
  \includegraphics*[width=7cm]{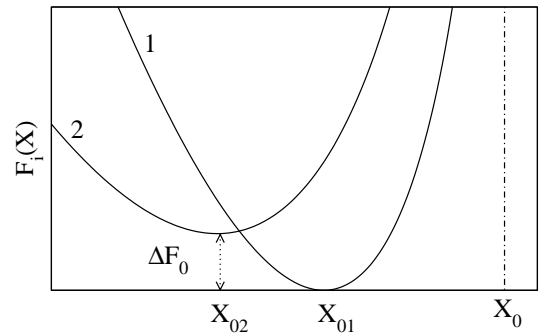}
  \caption{Free energy surfaces of ET $F_i(X)$ for the initial ($i=1$,
    ``1'') and final ($i=2$, ``2'') ET states.  Shown in the plot are
    the free energy minima $X_{0i}$, the free energy
    gap $\Delta F_0=F_{02}- F_{01}$, and the upper boundary $X_0$ for the
    reaction coordinate $X$. }
  \label{fig:2}
\end{figure}

The Gaussian integral in eq \ref{eq:2-7} can be taken exactly with the result
\begin{equation}
  \label{eq:2-8}
  F_i(X) = - \kappa_i X - \beta^{-1}\ln\left[\frac{\sinh \chi(X)}{\chi(X)}\right]
\end{equation}
where
\begin{equation}
  \label{eq:2-9}
     \chi(X) = 2\beta\sqrt{\kappa_i^3\lambda_i(X_{0} - X)}
\end{equation}
The parameters in eqs \ref{eq:2-8} and \ref{eq:2-9} are related to the
properties of the polarizable donor-acceptor complex and the polar solvent by the following
set of equations. The parameters $\kappa_i$ are  
\begin{equation}
  \label{eq:2-10}
  \kappa_i = \left( 2a_n f_{ni} \Delta \alpha \right)^{-1}
\end{equation}
The parameter $f_{ni}$ in eq \ref{eq:2-10} is responsible for the
enhancement of the effective solute dipole by the interaction of the
solute polarizability with the nuclear reaction field
(cf.\ to eqs \ref{eq:2-1} and \ref{eq:2-2})
\begin{equation}
  \label{eq:2-11}
  f_{ni} = (1 - 2a_n \alpha_i )^{-1}
\end{equation}
Further,
\begin{equation}
  \label{eq:2-12}
  X_0 = I_2^{\text{np}} - I_1^{\text{np}} - \Delta mF_m - \frac{1}{2}\Delta\alpha F_m^2 + \frac{(\Delta m + \Delta \alpha F_m)^2}{2\Delta\alpha} 
\end{equation}
is the boundary of the fluctuation band of the reaction coordinate $X$
(Figure \ref{fig:2}).  The requirement $X< X_0$ implicit in the
definition of $\chi(X)$ is eq \ref{eq:2-9} does not allow the energy gap
fluctuations to exceed $X_0$. Finally, $\lambda_i$ in eq \ref{eq:2-9} is the
solvent reorganization energy
\begin{equation}
  \label{eq:2-13}
   \lambda_i = a_n f_{ni} \left[\Delta m + \Delta\alpha(R_i + F)\right]^2
\end{equation}
where $\Delta m = m_2 - m_1$ and reaction field $R_i$ is
\begin{equation}
  \label{eq:2-14}
  R_i = 2a_nf_{ni} m_i
\end{equation}

The function $\chi(X)$ in eq \ref{eq:2-8} does not carry index specifying
the ET state because of the relation connecting the parameters of the
two ET states:
\begin{equation}
  \label{eq:2-13-1}
   \kappa_1^3 \lambda_1 = \kappa_2^3 \lambda_2
\end{equation}
In addition, the parameters $\kappa_1$ and $\kappa_2$ are related by the
additivity constant
\begin{equation}
  \label{eq:2-15}
  \kappa_1 - \kappa_2 = 1 
\end{equation}

\begin{figure}[htbp]
  \centering
  \includegraphics*[width=7cm]{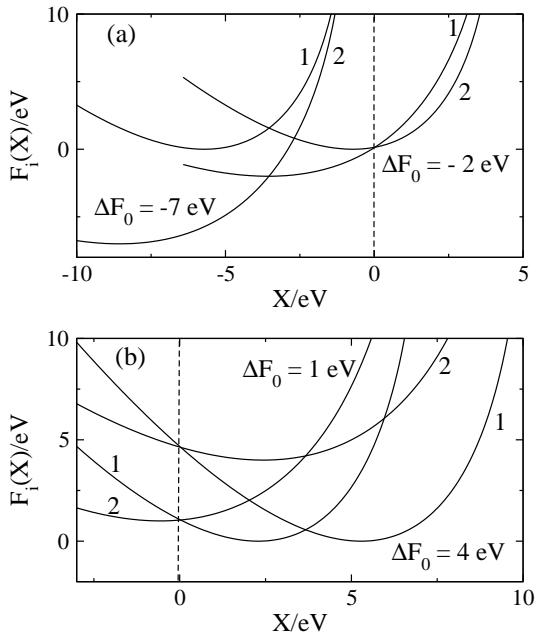}
  \caption{Free energy surfaces $F_i(X)$ at $\lambda_1=1$ eV and $\lambda_2=2$ eV.
    The free energy gap $\Delta F_0$ is: $-2$ eV and $-7$ eV in (a) and 1
    eV and 4 eV in (b). The free energy surfaces corresponding to $\Delta
    F_0=-7$ eV in (a) do not have the classical crossing point $X=0$
    indicated by the dashed lines in both panels. }
  \label{fig:3}
\end{figure}

The free energy surfaces defined in the range $X<X_0$ are
non-parabolic, as is depicted in Figures \ref{fig:2} and \ref{fig:3}.
Each of them passes through the minimum at $X=X_{0i}$ defined by the
condition
\begin{equation}
  \label{eq:2-16}
  d F_i(X)/ d X\bigg|_{X=X_{0i}} = 0 
\end{equation}
This equation can be reduced to the algebraic equation for $\chi(X)$
which always has a non-zero solution
\begin{equation}
  \label{eq:2-17}
  \chi\coth \chi - 1 = (2\beta\kappa_i^2\lambda_i)^{-1} \chi^2 
\end{equation}
The solution is simplified when $\chi_{0i}=\chi(X_{0i})\gg 1$. One then gets
\begin{equation}
  \label{eq:2-18}
  X_{0i} = X_0 - \kappa_i \lambda_i
\end{equation}
and the free energy at the minimum is
\begin{equation}
  \label{eq:2-19}
  F_{0i} = - \kappa_i X_0 - \kappa_i^2 \lambda_i
\end{equation}
In this limit, the second derivatives taken at the position of the
minima define the two solvent reorganization energies, as is the case
in standard formulations of ET theories
\begin{equation}
  \label{eq:2-20}
  \frac{d^2 F_i(X)}{  dX^2}\Bigg|_{X=X_{0i}} = \frac{1}{2\lambda_i} 
\end{equation}
Note that from eq \ref{eq:2-19} the fluctuation boundary $X_0$ is related to
the free energy gap $\Delta F_0 = F_{02}- F_{01}$ and two reorganization energies by
the relation
\begin{equation}
  \label{eq:2-20-1}
  X_0 = \Delta F_0 +\kappa_2^2\lambda_2 - \kappa_1^2 \lambda_1
\end{equation}

Taken together, eqs \ref{eq:2-8}--\ref{eq:2-20-1} provide an exact
model for the free energy surfaces of ET (Figure \ref{fig:2}) based on
three thermodynamic parameters, $\lambda_1$, $\lambda_2$, and $\Delta F_0$. The present
model thus extends the two-parameter Marcus-Hush theory, based on the
assumption $\lambda_1=\lambda_2$,\cite{Marcus:93} to three-parameters space.  In
view of the connection between $\kappa_1$ and $\kappa_2$ (eq \ref{eq:2-15}), one
of them can be considered as the non-parabolicity parameter:
\begin{equation}
  \label{eq:2-21}
  \kappa=\kappa_1 =\left(1 - \sqrt[3]{\frac{\lambda_1}{\lambda_2}} \right)^{-1}  
\end{equation}
In terms of the solute polarizability and the solvent nuclear response
function, the non-parabolicity parameter becomes
\begin{equation}
  \label{eq:2-22}
  \kappa = \frac{1 - 2 a_n \alpha_1}{2a_n\Delta\alpha} 
\end{equation}
Because of the choice $\Delta\alpha>0$ adopted here, $\lambda_1 < \lambda_2$ and $\kappa_i>0$.

It is easy to prove that $F_i(X)$ from eq \ref{eq:2-8} obey the energy
conservation requirement given by eq \ref{eq:1-1}.  Note that the
present model exhausts the space of thermodynamic parameters available
for the description of ET reactions. Only three thermodynamic
parameters, free energy gap between the minima and second cumulants of
the fluctuations around the minima, are allowed by the two-state
nature of the problem. Any extension of the theory for ET free energy
surfaces beyond the present level will require non-equilibrium
parameters to be involved.

The requirement of thermodynamic stability of the polarization
fluctuations limits the range of possible values of the reaction
coordinate, $X \leq X_0$. The reaction coordinate boundary $X_0$ is shown
by the dash-dotted line in Figure \ref{fig:2}. There is no
singularity of $F_i(X)$ at $X=X_0$ and the free energies $F_i(X)$ are
infinite at $X>X_0$.  For reaction coordinates sufficiently far from
$X_0$, the free energy surfaces can be conveniently re-written in the
form
\begin{equation}
  \label{eq:2-23}
  F_i(X) = F_{0i} + \kappa_i \left[\sqrt{X_{0i} + \kappa_i \lambda_i - X} - \sqrt{\kappa_i\lambda_i} \right]^2
\end{equation}
When the reorganization energies $\lambda_i$ are close to each other, the
non-parabolicity parameter $\kappa$ in eq \ref{eq:2-21} tends to infinity
and one obtains the standard Marcus-Hush parabolas by expanding eq 
\ref{eq:2-23} in $1/ \kappa$. From eq \ref{eq:2-23}, the activation
energy of ET is
\begin{equation}
  \label{eq:2-24}
  \Delta F_i = F_i(0) - F_{0i}=\kappa_i \left[\sqrt{X_{0i} + \kappa_i \lambda_i} - \sqrt{\kappa_i\lambda_i} \right]^2 
\end{equation}

\subsection{Qualitative results}
\label{sec:2-3}
\begin{figure}[htbp]
  \centering
  \includegraphics*[width=7cm]{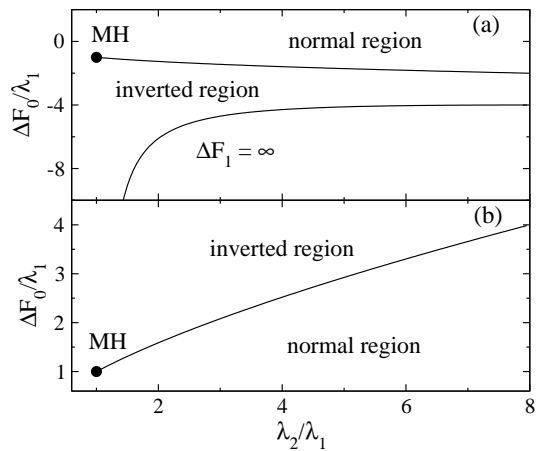}
  \caption{Normal and inverted region in the space of parameters $\Delta
    F_0/ \lambda_1$ and $\lambda_2 / \lambda_1$. Shown are the results for the forward
    reaction $1\to 2$ (a) and for the backward reaction $2\to 1$ (b). The
    quantum tunneling region with $\Delta F_1 = \infty$ in (a) is separated from
    the inverted region by the condition $X_0 = 0$, which puts the
    classical crossing point outside the range of reaction coordinates
    $X< X_0$ allowed by the condition of thermodynamic stability. The
    dots labeled as ``MH'' refer to the Marcus-Hush limit in which the
    transition from the normal to inverted region is given by the
    conditions: $\Delta F_0/ \lambda = - 1$ for $1\to 2$ and $\Delta F_0/ \lambda = 1$ for $2\to
    1$.  }
  \label{fig:4}
\end{figure}

The three-parameter model predicts some novel results regarding the
dependence of the reaction rates on the free energy gap (energy gap
law). In order to illustrate them, we will consider the transition
from the activated normal region to the activated inverted region
through the activationless transition for the forward reaction $1\to 2$
and the backward reaction $2\to1$ while maintaining $\lambda_1 < \lambda_2$ and the
definition of the free energy gap as $\Delta F_0 = F_{02}-F_{01}$ (Figure
\ref{fig:2}).  Our analysis here assumes $\chi_{0i}\gg1$, which holds for
most cases of interest. For the forward reaction $1\to 2$, the transition from
the normal to the inverted region is marked by the equation $X_{01}=0$,
which corresponds to a line in the space of parameters $\lambda_2/ \lambda_1 > 1$
and $\Delta F_0 / \lambda_1$ (Figure \ref{fig:4}):
\begin{equation}
  \label{eq:2-25}
  \Delta F_0 / \lambda_1 = \kappa(\kappa+1) - (\kappa -1)^2(\lambda_2/ \lambda_1) 
\end{equation}
The line shrinks into the point $\Delta F_0/ \lambda_1 = \pm 1$ in the Marcus-Hush
limit of equal reorganization energies (marked as ``MH'' in Figure 
\ref{fig:4}).

Lowering the free energy gap in the inverted region for the forward
reaction 1$\to 2$ leads to a new region of ET absent in the Marcus-Hush
theory. When the boundary of the band of allowed energy gaps $X_0$
crosses the transition state $X=0$, the classical crossing point of
two free energy surfaces falls outside the range of allowed reaction
coordinates. No crossing of free energy surfaces $F_i(X)$ is then
possible ($\Delta F_0 = -7$ eV in Figure \ref{fig:3}(a)), and the classical
reaction channel is closed, $\Delta F_1 = \infty$.  The reaction occurs only
through vibrational excitations of the final ET state 2 effectively
lowering the free energy gap. The transition to this new region, which
may be called ``quantum tunneling region'' ($\Delta F_1=\infty$ in Figure
\ref{fig:4}(a)), is marked by the line $X_0=0$ is the space of
parameters $\lambda_2/ \lambda_1 $ and $\Delta F_0 / \lambda_1$.

Due to the asymmetry of the free energy surfaces for $\lambda_2 / \lambda_1 > 1$,
the normal region spans different ranges of $\Delta F_0$ for positive and
negative free energy gaps. Figure \ref{fig:4} shows a broader
normal-range ET for $\Delta F_0 >0$.  This observation suggests that the
goal of bringing the exothermic recombination reaction to the normal
region (Figure \ref{fig:1}) is easier to achieve when the
reorganization energy of the charge-separated state D$^+$--A$^-$ is
much higher than the reorganization energy of the ground state D--A
(see Discussion).

\begin{figure}[htbp]
  \centering
  \includegraphics*[width=7cm]{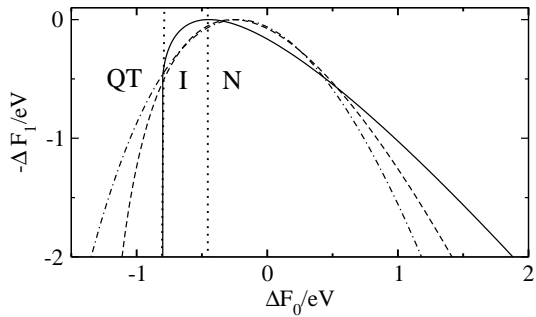}
  \caption{Energy gap law for the reaction 1$\to$2 ($\lambda_1 < \lambda_2$) at $\lambda_1=0.2$ eV
    and $\lambda_2=0.25$ eV (dash-dotted line), $\lambda_2= 0.4$ eV (dashed line),
    and $\lambda_2=2$ eV (solid line). The dotted lines separates the normal
    (N), inverted (I), and quantum tunneling (QT) regions for $\lambda_2=2$
    eV. }
  \label{fig:5}
\end{figure}

The overall classical energy gap law (no vibrational excitations) for
the present model is illustrated in Figure \ref{fig:5}. With
increasing the ratio of the reorganization energies $\lambda_2/ \lambda_1$, the
bell-shaped dependence of $-\Delta F_1$ on $\Delta F_0$ becomes increasingly
shallow in its right wing with positive energy gaps $\Delta F_0>0$,
approaching the linear energy gap law $\Delta F_1 \propto \Delta F_0$. On the other
hand, the approach of the fluctuation boundary $X_0$ to the transition
point $X=0$ squeezes the left wing narrowing the inverted region of
ET.

\begin{figure}[htbp]
  \centering
  \includegraphics*[width=7cm]{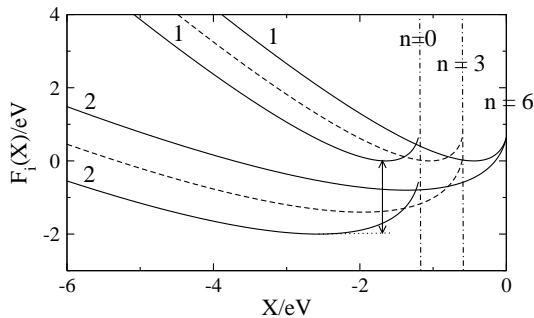}
  \caption{Inverted-region reaction $1\to2$ with $\lambda_1=0.2$ eV, $\lambda_2=1.0$ eV,
    and $\Delta F_0=-2$ eV (shown by the vertical arrow). Classical
    transitions between vibrationally ground-state surfaces are
    forbidden by the condition $X_0<0$. The dashed surfaces obtained
    for $n=3$ do not contribute to the Franck-Condon weighted density
    of states since $X_0 + n\hbar\omega_v<0$ for $n=3$. For $n\geq 6$, $X_0 +
    n\hbar\omega_v>0$ ($\hbar\omega_v=0.2$ eV) and these vibronic transitions contribute
    to the overall density of states. The dash-dotted lines indicate
    the fluctuation boundary $X_{0n}$ at $n=0$ and $n=3$; $X_{0n}=0$ at $n=6$. }
  \label{fig:6}
\end{figure}

The quantum vibronic excitations of the donor-acceptor complex can
be included in the standard way by considering the Franck-Condon weighted
density of states (FCWD) as a Poisson-weighted sum over the vibronic excitations
leading to $n$ vibrational quanta in the final ET state.\cite{BixonJortner:99} For
instance, for the forward reaction one gets
\begin{equation}
  \label{eq:2-26}
   \mathrm{FCWD}(X;1\to 2) = e^{-S}\sum_{n=0}^{\infty} \frac{S^n}{n!} e^{-\beta F_{1n}(X)} 
\end{equation}
Here, $S=\lambda_i/ \hbar\omega_v$ is the Huang-Rhys factor, $\omega_v$ is the characteristic vibrational
frequency, and $F_{1n}(X)$ is obtained from eq \ref{eq:2-8} by replacing there
$X_0$ with
\begin{equation}
  \label{eq:2-27}
  X_{0n}= X_0 + n\hbar\omega_v 
\end{equation}

The Arrhenius factor for the rate constant can be obtained (within a
factor) by putting $X=0$ in eq \ref{eq:2-26}.  Since the thermodynamic
stability of the nuclear fluctuations requires $X<X_{0n}$, only terms
corresponding to $X_{0n}>0$ in eq \ref{eq:2-27} will contribute to the
sum in eq \ref{eq:2-26}. This fact implies a modification of the
standard picture of the FCWD made of a Poisson-weighted sum of
vibronic transitions. Vibronic transitions with $n< -X_0/ (\hbar \omega_v)$
will not contribute to the FCWD as is illustrated by $n=0$ and $n=3$
in Figure \ref{fig:6}.  Only starting from vibrational excitation
number making $X_{0n}$ positive ($n=6$ in Figure \ref{fig:6}) will a
given vibronic transition contribute to the FCWD.  This fact will
result in a greater asymmetry of emission lines making the red-side
wing of an optical band more shallow than the blue-side wing.

\section{Electron transfer in ferroelectrics}
\label{sec:3}
Equation \ref{eq:2-13} for the solvent reorganization energies of a
polarizable donor-acceptor complex is the central result of the formal
theory relevant to our discussion of reorganization anisotropy in
ferroelectrics. Equation \ref{eq:2-13} predicts that the
polarizability change $\Delta \alpha$ couples to the macroscopic and local
(reaction) fields to modify the change in the solute dipole moment in
the course of electronic transition. Depending on the relative signs
of $\Delta m$ and $\Delta\alpha(R_i+F_m)$, the solute polarizability may increase or
decrease the solvent reorganization energy compared to the
non-polarizable limit.  The main question in this regard is whether
the term $\Delta\alpha(R_i + F_m)$ can achieve a magnitude comparable to $\Delta
m$.  Below, we address this question by MC simulations of a model
donor-acceptor diatomic in a dipolar ferroelectric solvent. Here, we
first provide some relevant estimates.

The mean-field Weiss theory of spontaneous dipolar
polarization\cite{Zhang:95} relates the macroscopic field in a sample
of aligned dipoles $m$ to their number density $\rho$ as
\begin{equation}
  \label{eq:3-6}
  F = (4\pi/3)\rho m 
\end{equation}
The term $\Delta\alpha F_m/m$ at $\rho^*=\rho\sigma^3=0.7$ is then $\simeq 3\Delta\alpha/ \sigma^3$ when the
solute dipole is aligned with the macroscopic field, $\sigma$ is the
solvent diameter.  The polarizability change may vary substantially
between different donor-acceptor units, but the ground-state
polarizability is close to $\sigma_0^3/16$ for many molecular systems
($\sigma_0$ is the effective solute diameter). If $\Delta\alpha$ is of the same order
of magnitude as the ground-state polarizability, $3\Delta\alpha/ \sigma^3$ scales as
$(\sigma_0/ \sigma )^3$.  For example, for primary charge separation in
\textit{Rhodobacter sphaeroids}, $\Delta m\simeq 53$ D.\cite{Bixon:95} If we
accept $\sigma=2.87$ \AA{} and $m=1.83$ D for water and $\Delta\alpha\simeq 800$
\AA,\cite{Middendorf:93} then $\Delta m/m\simeq 29$ and $\Delta\alpha F_m/m \simeq 100$ thus
resulting in comparable order-of-magnitude contributions from the
change in the permanent charges and from the interaction of the
polarizability change with the non-homogeneous electric field. In
addition, from the estimate of the local electric field of the protein
matrix at the position of the primary pair\cite{Steffen:94} $F\simeq 10^7$
V/cm, the induced dipole $\Delta \alpha F\simeq 27$ D is also comparable to $\Delta m \simeq
53$ D.

\subsection{Simulations}
\label{sec:3-1}
A fluid of soft spheres is known to spontaneously form a ferroelectric
liquid phase when conducting boundary conditions are employed in the
simulation protocol.\cite{Wei:92} This system was used here to model a
disordered phase with a molecular-scale macroscopic polarization
coupled to the electronic states of the donor-acceptor complex. Two
sets of simulations have been carried out. The fluid of soft dipolar
spheres was simulated in the first set in order to establish the range
of parameters for which ferroelectric phase can be detected. This was
followed by the second set of simulations in which polarizable
donor-acceptor complex was dissolved in the ferroelectric liquid.

MC simulations of the pure solvent employed NVT ensemble of $N=500$
particles in a cubic simulation cell at the reduced density of
$\rho^*=\rho\sigma^3=0.7$.  The molecules are interacting with the potential
\begin{equation}
  \label{eq:3-7}
  v(12) = 4 \epsilon \left(\sigma/r_{12}\right)^{12} - \mathbf{m}_1\cdot\mathbf{T}_{12}\cdot\mathbf{m}_2
\end{equation}
where $\epsilon$ is the repulsion energy, $\sigma$ is the diameter, $\mathbf{m}_j$
is the dipole moment with the magnitude $m$,
$\mathbf{T}_{12}=\nabla_1\nabla_2r_{12}^{-1}$ is the dipolar tensor, and
$r_{12}=|\mathbf{r}_1-\mathbf{r}_2|$. Simulations were done at
$\beta\epsilon=1.35$ and varying reduced dipole $(m^*)^2 = \beta m^2/ \sigma^3$. Periodic
boundary conditions and reaction-field correction for dipolar
interactions with conducting boundary conditions\cite{Allen:96} have
been employed. The length of simulations was $10^6$ cycles long far
from the transition point and up to $10^7$ cycles long close to
the transition to ferroelectric phase.

\begin{figure}[htbp]
  \centering
  \includegraphics*[width=7cm]{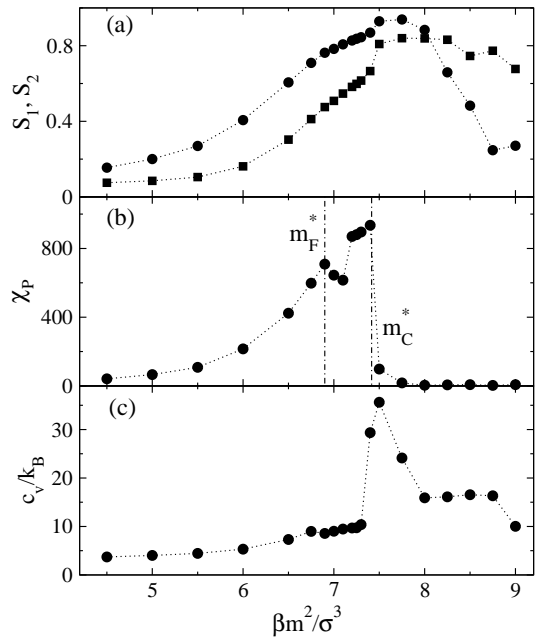}
  \caption{First ($S_1$, circles) and second ($S_2$, squares) order
    parameters of the fluid of dipolar soft spheres vs $\beta m^2/ \sigma^3$
    (a). Also shown are the dipolar susceptibility $\chi_P$ (b) and
    constant-volume heat capacity $c_V$ (c). The dotted lines connect
    the simulation points. The dash-dotted lines in (b) indicate the
    points of phase transition to ferroelectric fluid ($m_F^*$) and
    fcc crystal ($m_C^*$). }
  \label{fig:7}
\end{figure}

Transition to ferroelectric phase was monitored by calculating
the first-order, $S_1$, and second-order, $S_2$, parameters (Figure 
\ref{fig:7}(a)). The first-order (ferroelectric) parameter quantifies the
spontaneous polarization
\begin{equation}
  \label{eq:3-8}
  S_1 = M/Nm 
\end{equation}
where $\mathbf{M}$ is the total dipole moment of the solvent. The
second-order (nematic) parameter is defined as the largest eigenvalue
of the ordering matrix\cite{Allen:96}
\begin{equation}
  \label{eq:3-9}
  \mathbf{Q}=(2N)^{-1}\sum_j (3\mathbf{\hat e}_j\mathbf{\hat e}_j - \mathbf{1})
\end{equation}
where $\mathbf{\hat e}_j=\mathbf{m}_j/m$.  

Both order parameters change smoothly with $m^*$ and do not allow a
reliable identification of the transition point.\cite{Weis:05}
Susceptibilities, which are expected to show sharp spikes at the
points of phase transition,\cite{Binder:92} are better indicators.
Indeed, the dielectric susceptibility
\begin{equation}
  \label{eq:3-10}
  \chi_P = (\beta/V) \langle (\delta \mathbf{M})^2\rangle 
\end{equation}
shows the first peak at $(m_F^*)^2=6.9$ and the second peak at
$(m_C^*)^2= 7.5$ (Figure \ref{fig:7}(b)), where $V$ in eq
\ref{eq:3-10} is the solvent volume.  The heat capacity
\begin{equation}
  \label{eq:3-11}
   c_V = 3/2 + \beta^2 \langle (\delta E)^2 \rangle/N 
\end{equation}
on the contrary, shows only a little bump at $m^*_F$ and a strong peak
at $m^*_C$ (Figure \ref{fig:7}(c)), where $E$ in eq \ref{eq:3-11} is
the total energy of the fluid.  The order parameters also show weak
discontinuities at $m_C^*$. The examination of the density structure
factors reveals that the system is in the liquid state below $m^*_C$
and crystallizes into an fcc lattice at this value of the reduced
dipole.  The point $m^*_F$ corresponds to the transition to a
ferroelectric liquid.  Note that the value of $m^*_F$ obtained here is
very close to that predicted by a linear extrapolation of the phase
transition line recently reported by Weis\cite{Weis:05} for dipolar
hard spheres.

\begin{figure}[htbp]
  \centering
  \includegraphics*[width=7cm]{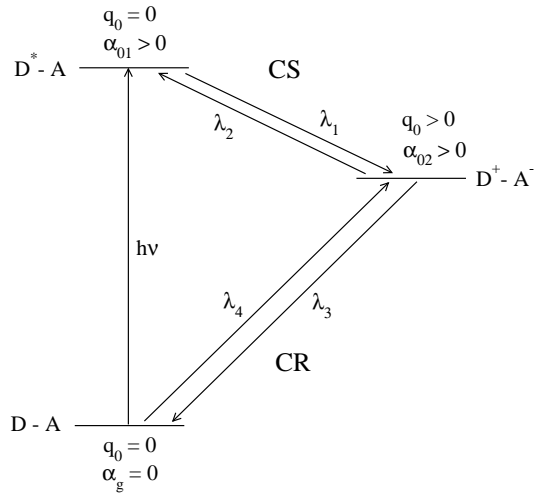}
  \caption{Charges and polarizabilities of the donor-acceptor complex
    in three states involved in photoinduced charge separation (CS) and
    charge recombination (CR). }
  \label{fig:8}
\end{figure}

The ferroelectric liquid at $(m^*)^2=7.0$, $\rho^*=0.7$, and $\beta\epsilon=1.35$
was used to study the dependence of the solvent reorganization energy
on solute polarizability.  In these simulations, a donor-acceptor
diatomic made of two hard fused spheres of diameters $\sigma_0/ \sigma = 1.5$
with the center-to-center separation $d/ \sigma = 0.6$ was inserted in the
center of a cubic simulation cell containing $N=500$ soft dipolar
particles at $\rho^*=0.7$.  Two opposite charges $q_0^*= \beta|q_0|/ \sigma = 10$
where placed at the centers of two spheres in the charge-separated
state thus producing the dipole moment $m_{02} = d |q_0|$. The induced
point dipole moment $\mathbf{p}_{0i} = \mathbf{\hat e}_0 \alpha_{0i} R_i$,
caused by the dipolar polarizability $\alpha_{0i}$ and reaction field
$R_i$, is placed at the midpoint of the line connecting the two
centers of the diatomic. $\mathbf{p}_{0i}$ is aligned along the unit
vector $\mathbf{\hat e}_0$ along the same line. The solute induced
dipole was equilibrated to each instantaneous configuration of the
solvent by iteration algorithm (more details on the simulation
protocol are given in ref \onlinecite{DMjpca:04}).  Ewald sums with
conducting boundary conditions\cite{Allen:96} were used to calculate
the solute-solvent electrostatic interactions.

Some complications arise from the fact that the director,
$\mathbf{\hat d}= \mathbf{M}/Nm$, fluctuates in the simulations. In
real liquid crystals, fluctuations of the macroscopic director occur
on the time-scale of microseconds to seconds,\cite{Vertogen:88} much
slower than orientational motions of molecular solutes.  The director,
therefore, does not fluctuate on the time-scale of the reaction. In
order to account for this hierarchy of relaxation times, the
simulation protocol was set up to keep the orientation of the
donor-acceptor complex fixed relative to the director.  The initial
configuration was created by inserting a sphere of diameter $\sigma_0$ into
the soft dipolar solvent followed by a short, $5\times10^3$ cycles,
equilibration run designed to establish the director $\mathbf{\hat
  d}$. The two spheres making the diatomic were then pulled apart at a
given angle to the director and this angle was adjusted after each
cycle over the $N$ solvent molecules. The data were collected from
simulation runs of the length $6\times 10^5-1.2\times 10^6$ cycles. In the
present simulations, the long axes of the diatomic was always aligned
with the director in such a way that the charge-transfer dipole $\Delta
\mathbf{m}_0$ is always parallel to the macroscopic field $\mathbf{F}$.

\begin{figure}[htbp]
  \centering
  \includegraphics*[width=7cm]{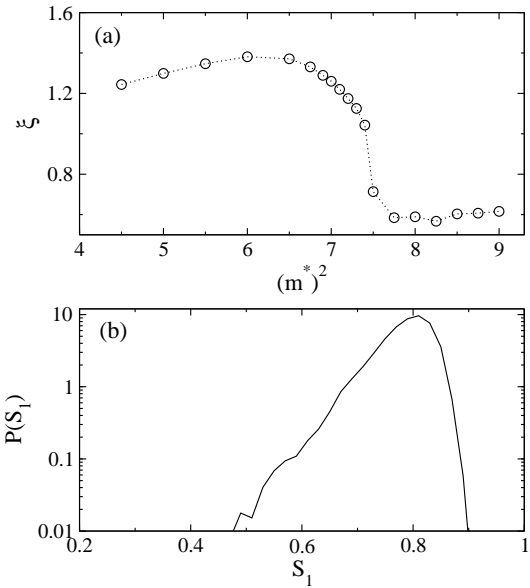}
  \caption{The linear response parameter $\xi$ from eq \ref{eq:3-15}
    (a) and the distribution function of the ferroelectric order
    parameter (b) obtained from MC simulations at $(m^*)^2=7.0$,
    $\beta\epsilon=1.35$, and $\rho^*=0.7$. The dotted line in (a) connects the
    simulation points. }
  \label{fig:9}
\end{figure}

\subsection{Energetics of photoinduced electron transfer}
\label{sec:3-2}
We have followed the basic design of photoinduced ET outlined in
Figure \ref{fig:1} and detailed in terms of charges and
polarizabilities in Figure \ref{fig:8}. It is assumed that the
donor-acceptor complex has no appreciable dipole moment and
polarizability in its ground state ($q_0=0$, $\alpha_g =0$).
Photoexcitation of the donor to state D$^*$--A lifts the
polarizability to $\alpha_{01} >0$ but does not substantially change the
charge distribution, $q_0=0$. Both the charge distribution and
polarizability change upon charge separation resulting in D$^+$--A$^-$
state ($q_0>0$, $\alpha_{02} > 0$). Because of the coupling of the
polarizability to the reaction field $R_i$ and the macroscopic
electric field $F_m$ (eq \ref{eq:2-13}), the reorganization energies
will differ for forward and backward electronic transitions for
both the charge separation and charge recombination steps of the
reaction mechanism. Therefore, we need four reorganization energies,
$\lambda_1$ and $\lambda_2$ for charge separation and $\lambda_3$ and $\lambda_4$ for charge
recombination (Figure \ref{fig:8}). 

\begin{table}[htbp]
  \centering
  \caption{Energetic parameters (eV) of the charge separation reaction (Figure \ref{fig:8}) 
    in which $\alpha_{01}$ is varied and $\alpha_{02}/ \sigma^3 = 0.2$ is kept constant; 
    $\beta q_0/ \sigma=10$ in the charge-separated state.}
  \begin{tabular}{ccccc}
\colrule
$\alpha_{01} / \sigma^3$ & $\lambda_1$\footnotemark[1] & $\lambda_2$ & $\Delta X_0$\footnotemark[2] 
           & $\Delta X_0$\footnotemark[3] \\
\colrule
0.2  & 0.324  & 0.456   &  0.817  &  0.770\\
0.3  & 0.344  &  0.371  &  0.750  &  0.715 \\
0.4  & 0.373  & 0.295   &  0.706  &  0.665 \\
0.5  & 0.477  &  0.229  &  0.704  & 0.666 \\
0.6  & 0.587  &  0.171  & 0.665   & 0.647 \\
0.7  & 0.761  &  0.123  & 0.612   & 0.640 \\
0.8  & 1.314  &  0.083  & 0.645   & 0.733 \\
\colrule
  \end{tabular}
\footnotetext[1]{Calculated from the simulation data assuming $\beta=40$ eV$^{-1}$. }
\footnotetext[2]{From the simulation data.}
\footnotetext[3]{From eq \ref{eq:3-14}. }
  \label{tab:1}
\end{table}

\begin{figure}[htbp]
  \centering
  \includegraphics*[width=7cm]{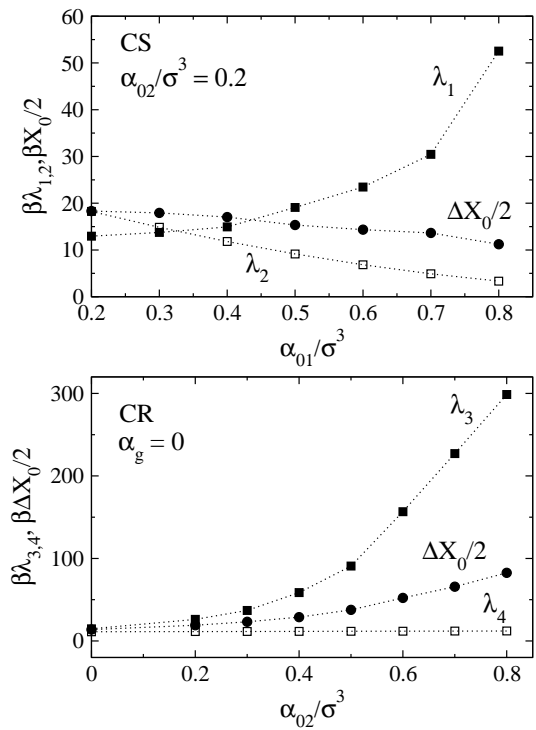}
  \caption{Reorganization energies and the Stokes shifts for the charge
    separation (CS) and charge recombination (CR) reactions (Figure
    \ref{fig:8}) vs polarizability of the photoexcited state $\alpha_{01}$
    (CS) and the polarizability of the charge-separated state
    $\alpha_{02}$ (CR); $q_0^*=10$, $\alpha_g=0$. }
  \label{fig:10}
\end{figure}

The reorganization energies were calculated from MC configurations as
second cumulants of the solute-solvent interaction
potential.\cite{DMjpca:04} The polarizability of the photoexcited state
$\alpha_{01}$ was varied at constant $\alpha_{02} /\sigma^3 =0.2$ for charge
separation and the polarizability of the charge-transfer state
$\alpha_{02}$ was varied at $\alpha_g=0$ for charge recombination.  Tables
\ref{tab:1} and \ref{tab:2} and Figure \ref{fig:10} summarize the
forward and backward reorganization energies for charge-separation and
charge-recombination reactions along with the Stokes shift
\begin{equation}
  \label{eq:3-13}
  \Delta X_0 = X_{01} - X_{02} 
\end{equation}
From eq \ref{eq:2-18}, the Stokes shift becomes
\begin{equation}
  \label{eq:3-14}
  \Delta X_0 = \kappa(\lambda_2-\lambda_1) - \lambda_2  
\end{equation}
The Stokes shift in eq \ref{eq:3-13} is equivalent to the difference
of absorption and emission maxima due to eq \ref{eq:1-1}. Equation 
\ref{eq:3-14} also applies to the Stokes shift defined in terms of
two first spectral moments
\begin{equation}
  \label{eq:2}
  \Delta X_0 = \langle X\rangle_1 - \langle X\rangle_2
\end{equation}
in the limit $\kappa_i\sqrt{\beta\lambda_i}\gg 1$.

The direct comparison of the results of simulations to the
three-parameter model discussed in Section \ref{sec:2} is complicated
by the narrow range of solvent dipoles for which ferroelectric phase
can be detected.  The close proximity of two phase transition points
makes the statistics of polarization fluctuations non-Gaussian, in
contrast to the Gaussian approximation adopted in eq \ref{eq:2-5}.
There are several indications of significant deviations from the
Gaussian statistics. The ratio
\begin{equation}
  \label{eq:3-15}
  \xi = - \langle v_{ss} \rangle / \beta \langle(\delta v_{ss})^2 \rangle  
\end{equation}
is equal to one ($\xi=1$) when polarization fluctuations are Gaussian
(linear response approximation).\cite{DMjcp1:99} Here, $\langle v_{ss}\rangle$ is
the average electrostatic interaction energy of a liquid particle with
the rest $N-1$ particles in the liquid and $ \langle(\delta v_{ss})^2 \rangle$ is the
variance of this electrostatic interaction. As is seen from Figure
\ref{fig:9}(a), this parameter is higher than one in the paraelectric
phase and drops sharply at the transition to the ferroelectric phase.
In addition, the distribution of macroscopic polarization $M$ in
the simulation box is non-Gaussian. A shoulder seen in Figure
\ref{fig:9}(b) points to the importance of a $\propto M^4$ term in the
polarization functional such as that present in the Landau theory of
phase transitions.\cite{Vertogen:88}

In terms of solvation properties, in the absence of polarizability
change of the solute, $\Delta X_0/2$ should be equal to $\lambda=\lambda_1=\lambda_2$. The
first lines in Tables \ref{tab:1} and \ref{tab:2} correspond to
exactly this situation. The difference between two reorganization
energies, as well as deviations of both of them from $\Delta X_0/2$ is
another indication of the non-Gaussian statistics of the solvent
polarization leading to non-linear solvation.  Despite these
complications, the direct calculations of the Stokes shift from
simulations compare semi-quantitatively to the results of applying eq
\ref{eq:3-14} (Tables \ref{tab:1} and \ref{tab:2}). The analytical
model developed in Section \ref{sec:2-2}, therefore, captures the
basic thermodynamics of ET with markedly different reorganization
energies.

\begin{table}[htbp]
  \centering
  \caption{Energetic parameters (eV) of the charge 
    recombination reaction (Figure \ref{fig:8}) for which $\alpha_{02}$ 
    is varied and $\alpha_g=0$ is kept constant; $\beta q_0/ \sigma=10$ in the charge-separated state. }
  \label{tab:2}
  \begin{tabular}{ccccc}
\colrule
$\alpha_{02}/ \sigma^3$ & $\lambda_3$\footnotemark[1] & $\lambda_4$ & $\Delta X_0 $\footnotemark[2] & $\Delta X_0$\footnotemark[3] \\
\colrule
0.0  & 0.369   &  0.279   & 0.649    &  0.642 \\
0.2  &   0.654 &  0.284  &   0.911  & 0.871 \\
0.3  &  1.115  &  0.287  &  1.115  &  1.051 \\
0.4  &  1.466  &   0.290 &  1.466  &  1.352 \\
0.5  &   2.270 &    0.293 &  2.270  &   1.728 \\
0.6  &  2.562  &   0.296 &   2.562  &   2.357 \\
0.7  & 3.240  &  0.299  &   3.240  &  2.928  \\
0.8  &   7.466  &  0.303 &   4.079 &   3.446 \\
\colrule
  \end{tabular}
\footnotetext[1]{Calculated from the simulation data assuming $\beta=40$ eV$^{-1}$.}
\footnotetext[2]{From the simulation data.}
\footnotetext[3]{From eq \ref{eq:3-14}. }
\end{table}

\section{Discussion}
\label{sec:4}
\begin{figure}[htbp]
  \centering
  \includegraphics*[width=7cm]{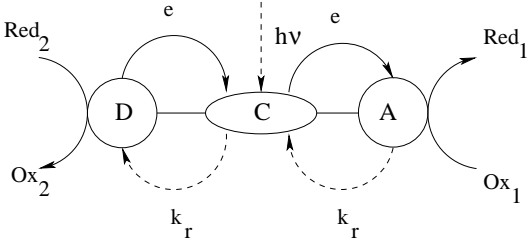}
  \caption{Reaction center model.\cite{Huynh:04,Acevedo:05} The unit C
    is photoexcited to obtain the electron from donor D and transfer it
    to acceptor A. The recombination rates $k_r$ at each stage should
    be sufficiently low to allow normally slow catalytic reduction
    (Ox$_1\to$Red$_1$) and oxidation (Red$_2\to$Ox$_2$) reactions to
    occur.  }
  \label{fig:11}
\end{figure}

The basic design of an artificial photosynthetic device, as advances
by Meyer and co-workers,\cite{Huynh:04,Acevedo:05} is shown in Figure
\ref{fig:11}. It anticipates the creation of ordered arrays where donors
and acceptors are connected to catalytic sites where high-energy
reactions can occur (e.g., splitting of water or reduction of carbon
dioxide to carbohydrates).  This design requires efficient separation
of the electron and the hole. Unless this separation is achieved by
high mobility of each charge in molecular arrays\cite{Hoertz:05} or in
valence and conduction bands of a semiconductor,\cite{Lewis:05} high
branching ratio between charge separation and charge recombination is
required at each step of productive charge transfer to facilitate the
redox reactions which are normally significantly slower than ET steps.

The model presented here offers some novel opportunities in terms of
varying the parameters of ET reactions and activation barriers.  The
model emphasizes the coupling of the solute polarizability change,
achieved in the course of electronic transition, to the 
electric field at the position of the donor-acceptor complex. The
electric field has two components, the reaction field of the polar
environment $R_i$ and the macroscopic field $F_m$. The coupling of the
overall field $R_i+F_m$ to the polarizability change creates the effective
dipole moment change of the donor-acceptor complex (eq \ref{eq:2-13})
\begin{equation}
  \label{eq:4-1}
  \Delta m_{\text{eff},i} = \Delta m + \Delta \alpha (R_i + F_m)
\end{equation}
The reaction field $R_i$, which depends on the electronic state of the
donor-acceptor complex, is responsible for the distinction between the
forward and backward reorganization energies, $\lambda_1 \neq \lambda_2$ and $\lambda_3\neq
\lambda_4$ as is shown in Figure \ref{fig:10} and listed in Tables
\ref{tab:1} and \ref{tab:2}. Note that the asymmetry in the reorganization
energies up to a factor of 25 obtained in our simulations (Table
\ref{tab:2}) is the largest ever observed in either computer or
laboratory experiment.

\begin{figure}[htbp]
  \centering
  \includegraphics*[width=7cm]{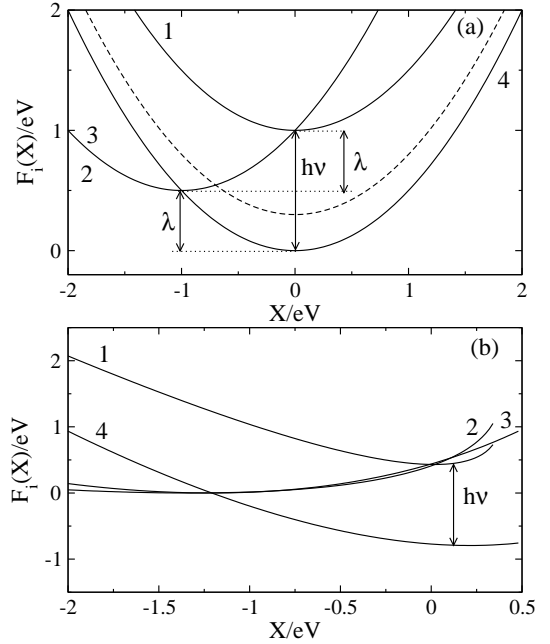}
  \caption{Free energy surfaces of photoinduced ET in the Marcus-Hush
    picture (a) and in the present model with different reorganization
    energies (b).  The free energy surfaces are plotted against the
    reaction coordinate corresponding to the energy gap between the
    charge-separated and photoexcited states. The parameters are
    chosen to show the activationless pathway from the photoexcited
    state 1 to the charge-separated state 2 and from 2 to the ground
    state 3.  The free energy surfaces 2 and 3 coincide in the
    Marcus-Hush picture, but are distinct in the present model. In (a)
    the dashed line indicates the free energy surface for the
    normal-region charge recombination when energetic efficiency is
    below 50\%; $h\nu$ indicates the energy of photoexcitation. The
    reorganization energies are: $\lambda=1$ eV (a) and $\lambda_1=0.2$ eV,
    $\lambda_2=2.0$ eV, $\lambda_3=1.0$ eV, and $\lambda_4=0.5$ eV (b). }
  \label{fig:12}
\end{figure}

A note on the dual nature of the reorganization energy is relevant
here.  The reorganization energy for each ET state is calculated on
the configurations of the solvent in equilibrium with the
donor-acceptor complex in that given state. Therefore, the
reorganization energy carries information about a particular
electronic state of the donor-acceptor complex. On the other hand, the
value that is being averaged is the change in the interaction
potential, which is a characteristic of a given transition. As a
result of this duality, the electronic charge-separated state of the
donor-acceptor complex D$^+$--A$^-$ can be characterized by two,
generally unequal, reorganization energies $\lambda_2$ and $\lambda_3$ reflecting
transitions to two different electronic states, photoexcited and
ground.

In the present picture, all reorganization energies $\lambda_1$, $\lambda_2$,
$\lambda_3$, and $\lambda_4$ are allowed to be different, producing reach
energetics of photoinduced charge separation and charge recombination.
The difference in scenarios for photoinduced ET in the Marcus-Hush
model and in the present formulation is illustrated in Figure
\ref{fig:12}.  The plots show the free energy surfaces vs the reaction
coordinate corresponding to the energy gap $X$ between the
charge-separated and photoexcited states. The energy gap for the
charge recombination reaction is then $-\Delta F_{\text{CS}} - \Delta
F_{\text{CR}} + X$, where $\Delta F_{\text{CS,CR}}$ are the free energy
gaps for charge-separation and charge-recombination reactions.  In the
Marcus-Hush model, activationless charge separation and charge
recombination is achieved at the photoexcitation energy $h\nu = 2 \lambda$
(Figure \ref{fig:12}(a)).  Therefore, the desire to move the
recombination reaction to the normal region requires energetic
efficiency of the photosynthetic apparatus to be below 50\% (dashed
line in Figure \ref{fig:12}(a)).  This situation changes when
different reorganization energies are allowed in the construction of
the free energy surfaces.  The plots in Figure \ref{fig:12}(b) are
made with $\lambda_2/\lambda_1=10$ and $\lambda_3/ \lambda_4 = 5$.  When activationless
reactions are realized for $1\to 2$ and $2\to3$ transitions, the asymmetry
of the energy gap law leads to a higher energetic efficiency of 70\%.
The disadvantage of this scheme is a shallow shape of the
charge-separated free energy surface making activation energy for
charge recombination rather weakly dependent on changes in the free
energy gap.

The role of the reaction field factor in eq \ref{eq:4-1} will diminish
in weakly polar media characteristic of protein electron transfer. The
effect of a strong local electric field $F_m$ then becomes more
important. If the term $\Delta \alpha R_i$ is small compared to $\Delta m$, we can
simplify the effective dipole moment change to the form
\begin{equation}
  \label{eq:4-2}
  \Delta m_{\text{eff}} = \Delta m + \Delta \alpha F_m
\end{equation}
Since $\Delta m_{\text{eff}}$ does not depend on the electronic state, the
forward and backward reorganization energies are the same ($\lambda_1=\lambda_2$,
$\lambda_3=\lambda_4$), according to the Marcus-Hush theory. However, the factor
$\Delta \alpha F_m$ can be responsible for the difference in the reorganization
energies between charge separation and charge recombination reactions.

\begin{figure}[htbp]
  \centering
  \includegraphics*[width=7cm]{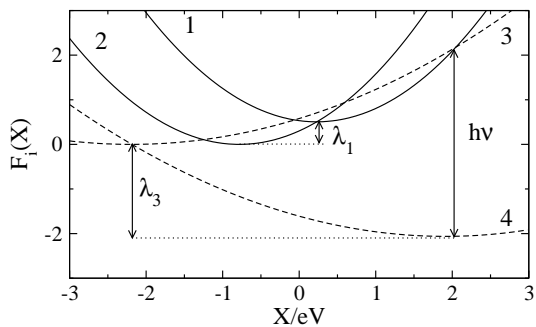}
  \caption{Free energy surfaces in the Marcus-Hush picture when 
           the reorganization energy for charge recombination  $\lambda_3=\lambda_4=2.0$ eV  
           is higher than reorganization energy for charge separation $\lambda_1=\lambda_2=0.5$ eV  
           due to the coupling of the polarizability change to the macroscopic electric field
           (eqs \ref{eq:2-13} and \ref{eq:4-2}). }
  \label{fig:13}
\end{figure}

Imagine a situation in which the photoexcited state is significantly
more polarizable than the ground state and the charge-separated state
has about the same polarizability as the photoexcited state,
$\Delta\alpha_{\text{CS}}\simeq 0$. Then, in weakly polar solvents, $\lambda_1\simeq\lambda_2$ and
both reorganization energies are small because of the low polarity of
the medium. For the charge recombination reaction, $\Delta m < 0 $ and
$\Delta\alpha_{\text{CR}} < 0$. The permanent dipole $\Delta m$ and the induced
dipole $\Delta \alpha F_m$ in eq \ref{eq:4-2} add up when $F_m>0$ and subtract
when $F_m<0$. In the former case, the charge-recombination
reorganization energy $\lambda_3\simeq\lambda_4$ is higher than the charge-separation
reorganization energy $\lambda_1\simeq\lambda_2$.  Note that this picture does not
contradict to the energy conservation requirement given by eq
\ref{eq:1-1} since the photoexcited state D$^*$--A and the ground
state D--A are two different electronic states characterized by
different polarizabilities.

The normal-region ET can be realized within the Marcus-Hush picture
when the reorganization energy is greater than the
charge-recombination free energy gap $\lambda_3 > \Delta F_{\text{CR}}$ (Figure
\ref{fig:13}). As in Figure \ref{fig:12}(a), the energetic efficiency
is about 50\%, and it drops when charge recombination is shifted into
the normal region. However, even in this limit of reduced flexibility
of altering the ET parameters, the presence of the macroscopic
electric field in the expression for the reorganization energy (eqs
\ref{eq:2-13} and \ref{eq:4-2}) opens the door to manipulations of
reorganization parameters through proper design of mutual signs of the
polarizability change and the direction of the macroscopic field.  

We can conclude that the combination of a highly polarizable
donor-acceptor complexe with macroscopic electric field creates a
principal possibility for suppressing the recombination reaction in
photosynthetic ET. From a more general perspective, the notion of
non-equal reorganization energies and non-parabolic free energy
surfaces, compliant with the condition of energy conservation (eq
\ref{eq:1-1}), opens the door to new models of ET activation which may
eventually lead to a practical solution of the photosynthesis
problem.

\begin{acknowledgments}
  This research was supported by the National Science Foundation
  (CHE-0304694). 
\end{acknowledgments}  

\bibliographystyle{achemso}
\bibliography{/home/dmitry/p/bib/chem_abbr,/home/dmitry/p/bib/photosynth,/home/dmitry/p/bib/et,/home/dmitry/p/bib/liquids,/home/dmitry/p/bib/solvation,/home/dmitry/p/bib/etnonlin,/home/dmitry/p/bib/dm,/home/dmitry/p/bib/dynamics,/home/dmitry/p/bib/ferro,/home/dmitry/p/bib/lc}

\end{document}